\documentclass[letter]{aa}  
\usepackage{natbib}
\usepackage{xcolor}
\usepackage{graphicx}
\usepackage{natbib}
\bibpunct{(}{)}{;}{a}{}{,}
\usepackage{diagbox}
\usepackage{txfonts}
\usepackage{xcolor}
\newcommand{\blos}{$B_{\parallel}$}

\newcommand{\Av}{$A_{{V}}$}
\newcommand{\bperp}{$B_{\perp}$}
\usepackage{scalerel}

\newcommand{\HI}{H\protect\scaleto{$I$}{1.2ex}}
\newcommand{\vcoHi}{$v_{\text{CO - H\protect\scaleto{$I$}{1.2ex}}}$}
\newcommand{\vco}{$v_{\text{CO, LSR}}$}
\newcommand{\vhi}{$v_{\text{H\protect\scaleto{$I$}{1.2ex}, LSR}}$}

\newcommand{\kms}{km\,s$^{-1}$}

\usepackage{blindtext}
\usepackage{makecell}

\begin{document} 
  \title{Orion A's complete 3D magnetic field morphology}
  \author{ M. Tahani \inst{1}
        \and J. Glover\inst{1, 2}
          \and W. Lupypciw\inst{1,3}
          \and J.L. West\inst{4}
          \and R. Kothes\inst{1}
          \and R. Plume\inst{3}
          \and S. Inutsuka\inst{5}
          \and M-Y. Lee\inst{6}
          \and I.A. Grenier\inst{7}
          \and L.B.G. Knee\inst{8}
          \and J.C. Brown\inst{3}
          \and Y. Doi\inst{9}
          \and T. Robishaw\inst{1}
          \and M. Haverkorn\inst{10}
         }

   \institute{Dominion Radio Astrophysical Observatory, Herzberg Astronomy and Astrophysics Research Centre, National Research Council Canada, P. O. Box 248, Penticton, BC V2A 6J9 Canada\\
   \email{mehrnoosh.tahani@nrc.ca}
   \and Physics \& Astronomy, University of Victoria, Victoria, British Columbia, Canada
   \and Physics \& Astronomy, University of Calgary, Calgary, Alberta, Canada
    \and
    Dunlap Institute for Astronomy and Astrophysics University of Toronto, Toronto, ON M5S 3H4, Canada
    \and 
    Department of Physics, Graduate School of Science, Nagoya University, Furo-cho, Chikusa-ku, Nagoya 464-8602, Japan
    \and
    Korea Astronomy and Space Science Institute, 776 Daedeok-daero, 34055 Daejeon, Republic of Korea 
    \and
    Universit\'{e} de Paris and Universit\'{e} Paris Saclay, CEA, CNRS, AIM, F-91190 Gif-sur-Yvette, France
    \and 
    Herzberg Astronomy and Astrophysics Research Centre, National Research Council Canada, 5071 West Saanich Road, Victoria BC V9E 2E7, Canada
    \and 
    Department of Earth Science and Astronomy, Graduate School of Arts and Sciences, The University of Tokyo, 3-8-1 Komaba, Meguro, Tokyo 153-8902, Japan
    \and 
    Department of Astrophysics/IMAPP, Radboud University, P.O. Box 9010,
6500 GL Nijmegen, The Netherlands
             }

   \date{Received Date; accepted Date}

\titlerunning{Orion A 3D magnetic field}
\authorrunning{M. Tahani et al.}
 
\abstract
{Magnetic fields permeate the interstellar medium and are important in the star formation process. Determining the 3D magnetic fields of molecular clouds will allow us to better understand their role in the evolution of these clouds and formation of stars. We fully reconstruct the approximate three-dimensional (3D) magnetic field morphology of the Orion A molecular cloud (on scales of a few to $\sim 100$\,pc) using Galactic magnetic field models, as well as present line-of-sight and plane-of-sky magnetic field observations. While previous studies identified the 3D magnetic field morphology of the Orion A cloud as an arc shape, in this study we provide the orientation of this arc-shaped field and its plane-of-sky direction, for the first time. We find that this 3D field is a tilted, semi-convex (from our point of view) structure and  mostly points in the direction of decreasing latitude and longitude on the plane of the sky, from our vantage point. The previously identified bubbles and events in this region were key in shaping this arc-shaped magnetic field morphology. }

   
\maketitle
%
%-------------------------------------------------------------------
\section{Introduction}
\label{introduction}

The role of magnetic fields on different scales of star formation (from the formation of clouds to the formation of stars) is poorly understood. A major limitation in this understanding is the lack of knowledge of the three-dimensional (3D) magnetic fields in the interstellar medium (ISM). Multi-wavelength magnetic field observations are necessary to determine the 3D magnetic fields associated with star-forming regions.

Far-infrared dust polarized emission~\citep[e.g.,][]{Houdeetal2004, Poidevinetal2011, Fisseletal2016, PattleFissel2019, Doietal2020} and near-infrared/optical ~\citep[e.g.,][]{PereyraMagalhaes2004, Clemensetal2020} starlight polarization observations have enabled us to probe the plane-of-sky orientation of magnetic fields (\bperp ) in a number of molecular clouds. Dust polarization observations revealed that magnetic field lines tend to be perpendicular to high column density ($> 10^{21.7}$\,cm$^{-2}$) filamentary structures~\citep{PlanckXXXII,  PlanckXXXV}, allowing for greater mass accumulations and denser filaments \citep[e.g.,][]{Inoueetal2018, HennebelleInutsuka2019}.

Radio observations of Faraday rotation~\citep{Tahanietal2018} and Zeeman measurements~\citep[e.g.,][]{Goodman1989, Heiles1997, Trolandetal2008} have been used to probe the line-of-sight magnetic fields (\blos ) of molecular clouds. \cite{Tahanietal2018} developed a new technique based on Faraday rotation measurements for determining the strength and direction of \blos\ associated with molecular clouds\footnote{\citet{Tahanietal2018} code for determining \blos\ is available at \url{https://github.com/MehrnooshTahani/MappingBLOS_MolecularClouds}}. In this technique they incorporated an approach based on relative measurements to estimate the amount of rotation measure (RM) caused by molecular clouds, using RM data from \citet[][]{Tayloretal2009}. To extract the magnetic field strengths from the RMs, they estimated the electron column density of the molecular cloud at the position of each RM point using a chemical evolution code \citep{Gibsonetal2009} and extinction maps for each cloud. 

\begin{figure}[htbp]
\centering
\includegraphics[scale=0.3, trim={0cm 0cm 0cm 0cm},clip]{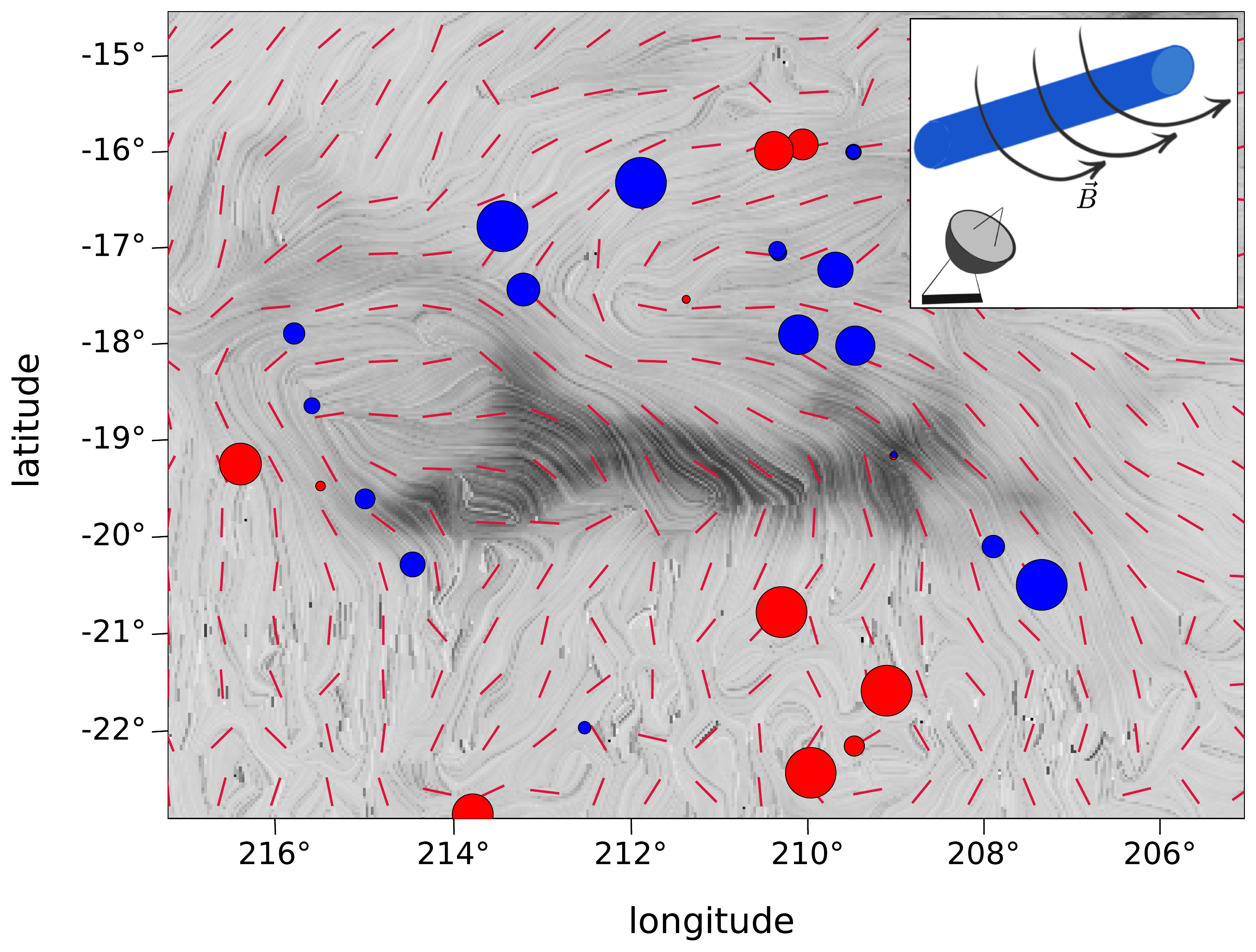}
\caption{The observed magnetic fields of the Orion A molecular cloud. The \blos\ data were obtained by~\citet{Tahanietal2018} and the \bperp\ lines were observed by the Planck Space Observatory. The blue (red) circles show magnetic fields toward (away from) us, and the size of the circles represents the strength of the magnetic fields. The red and drapery lines show the \bperp orientation. The background gray-scale image is the visual extinction map obtained by \citet{Kainulainenetal2009}. \cite{Tahanietal2019} suggested that an arc-shaped magnetic field morphology (as illustrated in the inset cartoon) is the most probable model that can explain the observed \blos\ reversal across the cloud.
} 
\label{fig:OrionBow}
\end{figure}

The observations of \bperp\ and \blos\ are slowly paving the way for the determination of 3D magnetic fields of molecular clouds. \citet{Tahanietal2019} investigated the 3D magnetic field morphology of the Orion A molecular cloud using both \blos\ \citep{Tahanietal2018} and \bperp\ \citep{PlanckXXXV} observations. The line-of-sight magnetic field observations in this region showed that the \blos\ reverses direction from one side of this filamentary-shaped cloud to the other (perpendicular to the filament axis; illustrated in Fig.~\ref{fig:OrionBow}). \citet{Tahanietal2019} constructed models to account for this reversal and compared their synthetic observations to the observed \bperp\ and \blos . Using Monte-Carlo simulations and chi-squared probability values (as well as examining systematic biases between the two observing techniques), they concluded that an arc-shaped\footnote{Also referred to as bow-shaped; pronounced /b\={o}/ as in rainbow or bow and arrow} magnetic field morphology was the most likely magnetic structure for the Orion A molecular cloud. 

This arc-shaped morphology is illustrated in the inset of Fig.~\ref{fig:OrionBow}. The background gray-scale image in this figure depicts the visual extinction map of Orion A \citep[in units of magnitude of visual extinction or \Av ;][]{Kainulainenetal2009}, and the red and drapery lines~\citep[made using the line integration convolution technique\footnote{\url{https://pypi.org/project/licpy/}},][]{Cabral1993LIC} show the \bperp\ observed by the Planck Space Observatory, with blue (red) circles indicating magnetic fields toward (away from) us. The magnetic field on the Galactic north side of the cloud points toward us, while on the Galactic south side it points away from us. This \blos\ reversal was previously observed using Zeeman measurements in the study of~\citet{Heiles1997}, which predicted an arc-shaped magnetic field morphology caused by recurrent shocks from nearby supernovae in the Orion-Eridanus superbubble.

%----------------------------------------------------
%------------------------------------------

\begin{figure}[htbp]
\centering
\includegraphics[scale=0.28, trim={1cm 0cm 3cm 1cm},clip]{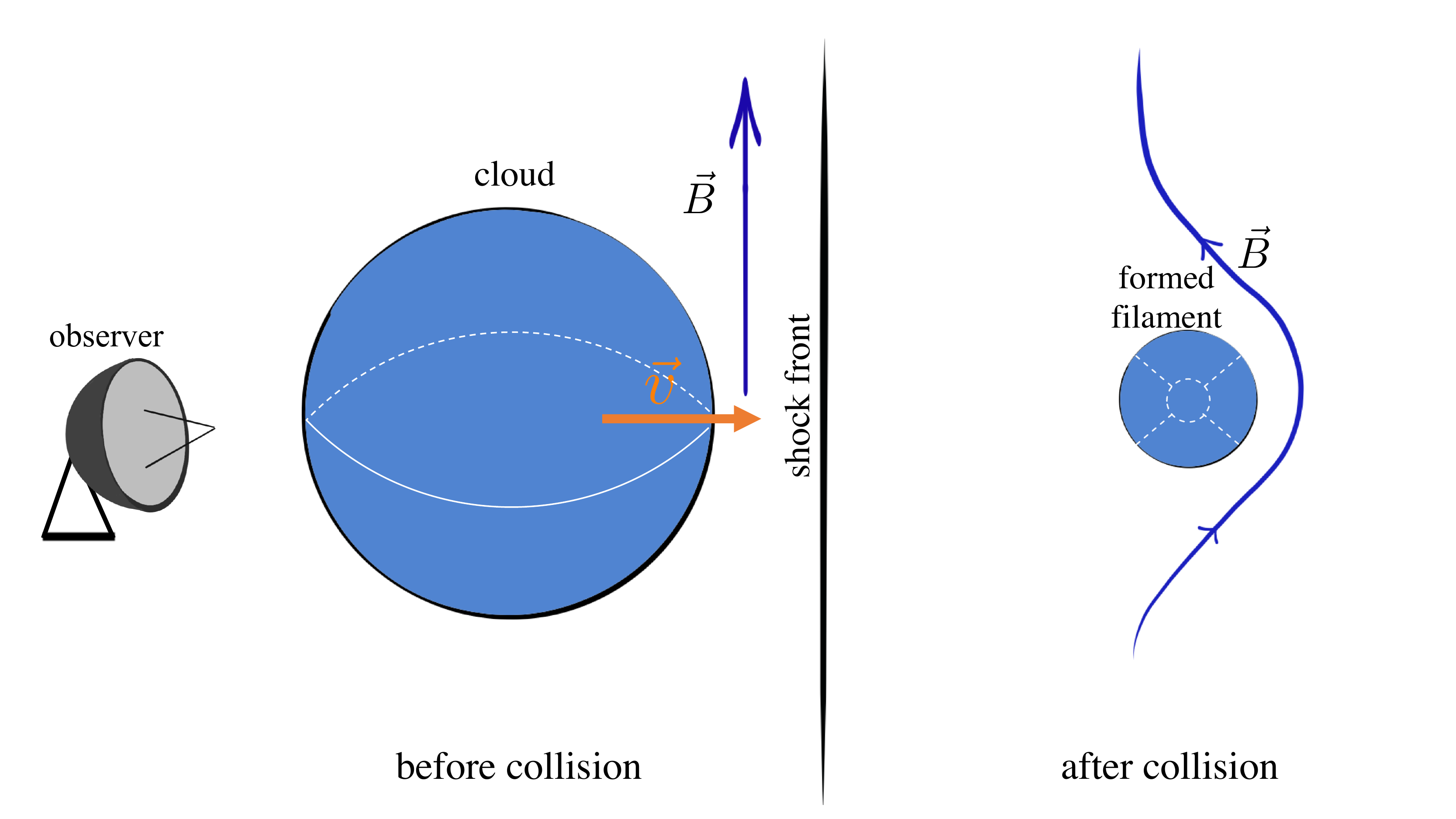}
\caption{Formation of an arc-shaped magnetic field morphology around filamentary molecular clouds as described by \cite{Inoueetal2018}. After the interaction between the cloud and shock-front, with the velocity \vec{v} in the co-moving frame of the shock-front, a filamentary structure is formed (shown in an end-on view in the right side). \vec{B} shows the direction of the initial magnetic field before the collision and the morphology of the magnetic field after the collision.}
\label{fig:BLOSDirection}
\end{figure}

An arc-shaped magnetic field morphology has also been generated in the ideal magnetohydrodynamic simulations of \citet{Inoueetal2018} and \citet{FukuiInoue2013}, and predicted in the \citet{Inutsukaetal2015} molecular-cloud-formation model \citep[also see][]{Inutsuka2016IAU}. Multiple compressions caused by expanding ISM bubbles are required for the formation of filamentary molecular clouds in the \citet{Inutsukaetal2015} model, where magnetic field lines with a component perpendicular to the direction of bubble expansion or propagation can be bent. 
Figure~\ref{fig:BLOSDirection} depicts a schematic cartoon of this interaction, which occurs between a relatively dense cloud ($\sim10^3$\,cm$^{-3}$) and a shock wave propagating in low density gas (\HI ) and bends the field lines around the formed filament. We refer to this cloud-formation model as the shock-cloud interaction (SCI) model. 
 
 %----------------------------------------------------
%----------------------------------------------------

While previous studies described the magnetic field morphology of the Orion A cloud as having an arc shape, its complete 3D orientation and direction (particularly when projected onto the plane of the sky) remained unknown, even in the 3D magnetic field study by~\citet{Tahanietal2019}. 
In this study, the complete 3D magnetic field morphology of the Orion A molecular cloud, including its direction and bending orientation, is reconstructed from data. To this end, we employ Galactic magnetic field (GMF) models, the \blos\ data of \cite{Tahanietal2018}, and the \bperp\ observed by the Planck Space Observatory (whereas in \citet{Tahanietal2019} only \bperp\ and \blos\ observations were used).  We present our approach and results in Sect.~\ref{Sec:reconstructed3DField}, discuss the role of surrounding structures, bubbles, and events in formation of this 3D field in Sect.~\ref{Sec:ArcFormation}, and provide a summary and conclusion in Sect.~\ref{Sec:summary}. Supplemental material, including the data used in this study (see Appendix~\ref{Sec:Appdata}), are provided in the appendix sections.  

\section{Results: 3D magnetic field morphology of Orion A}
\label{Sec:reconstructed3DField}

\begin{figure}[htbp]
\centering
\includegraphics[scale=0.3, trim={0cm 1.5cm 0.5cm 2.8cm},clip]{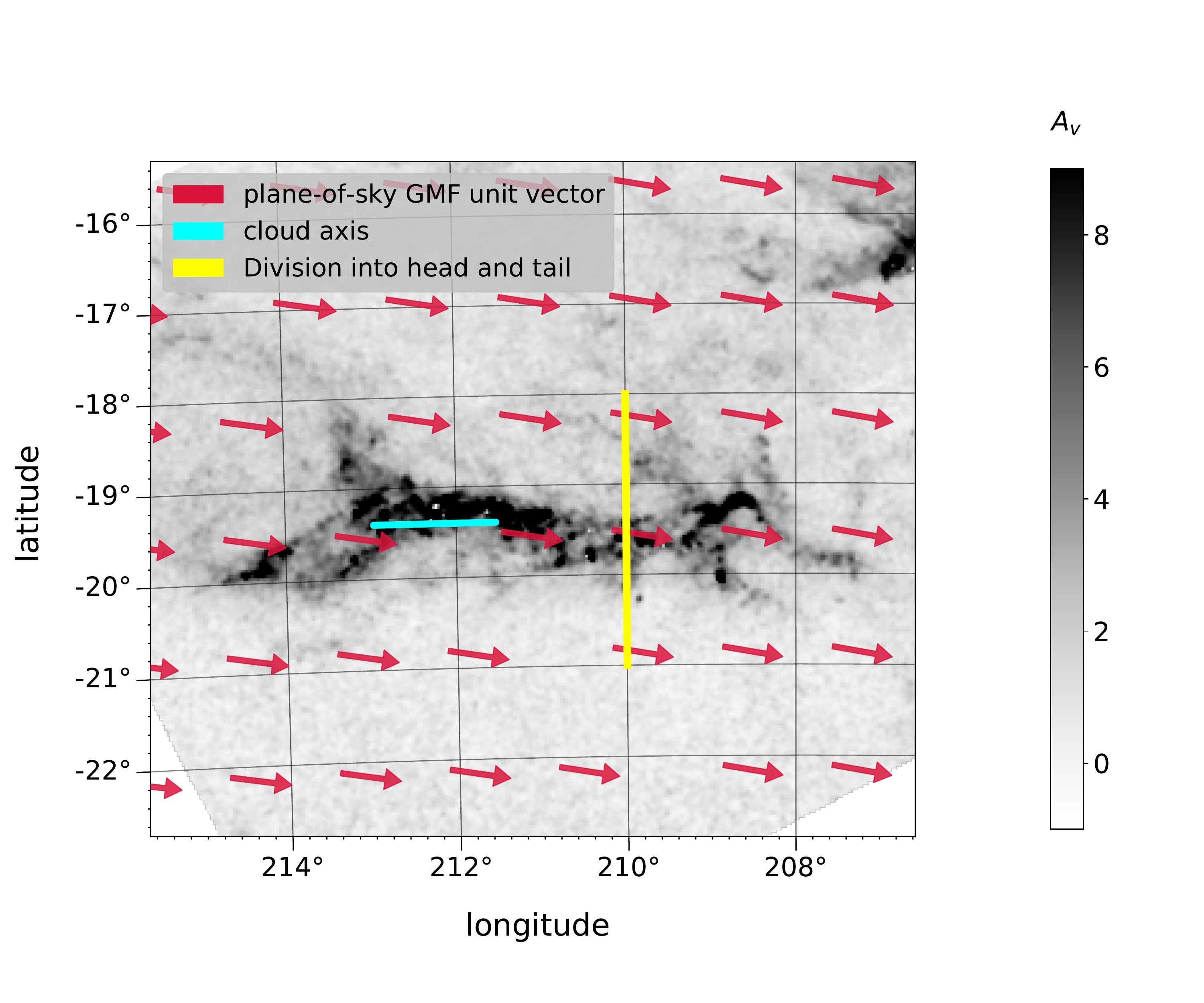}
\caption{Galactic magnetic field associated with the Orion A cloud. The background gray-scale image shows the extinction map from~\cite{Kainulainenetal2009}. The yellow line represents the division between the tail and the head of the cloud.  
The region of Orion A with longitudes less than $\sim 210^{\circ}$ is referred to as the ``head'', while the portion with longitudes greater than $\sim 210^{\circ}$ is referred to as the ``tail'' of Orion A. 
The red vectors show the Coherent GMF model projected onto the plane of the sky. The cyan line illustrates an approximate filament axis. }
\label{fig:GMFOrionA}
\end{figure}

We recreate the  3D magnetic field shape of the Orion A molecular cloud using GMF vectors as initial magnetic fields, the orientation of \blos\ reversal, and \bperp\ morphology (under the assumption of an arc-shaped magnetic morphology). To approximate the direction of the initial magnetic fields, we use the GMF model of \citet[][]{JanssonandFarrar2012}, neglecting the isotropic random field components (caused by ISM turbulence). We refer to this structure as the ``Coherent GMF'' model. Figure~\ref{fig:GMFOrionA} illustrates these GMF vectors projected onto the plane of the sky (red arrows) at the location of Orion A. 

\begin{figure}[htbp]
\centering
\includegraphics[scale=0.45, trim={0cm 0cm 0.cm 0.5cm},clip]{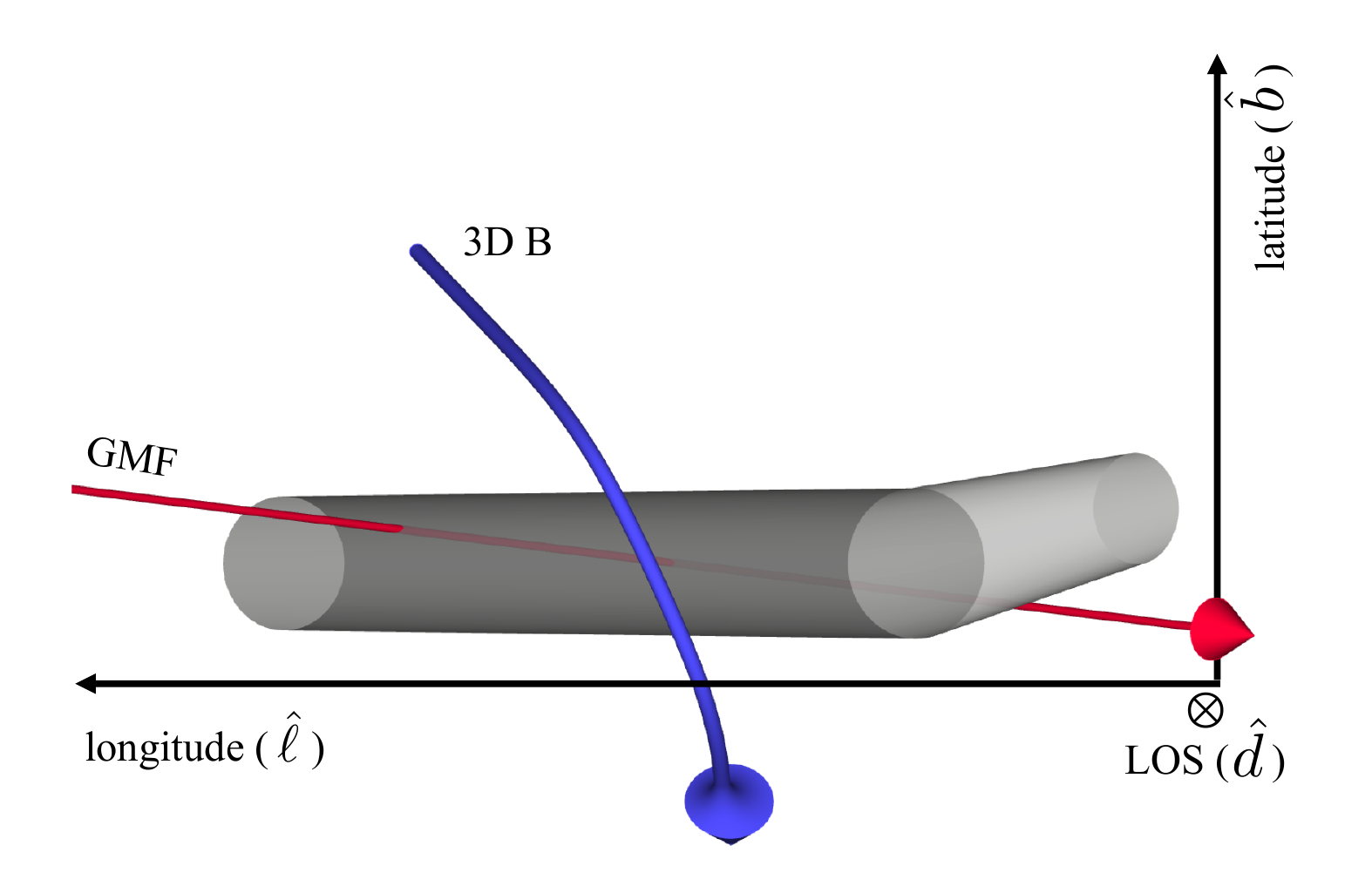}\\
\includegraphics[scale=0.45, trim={0cm 0cm 0.cm 0.cm},clip]{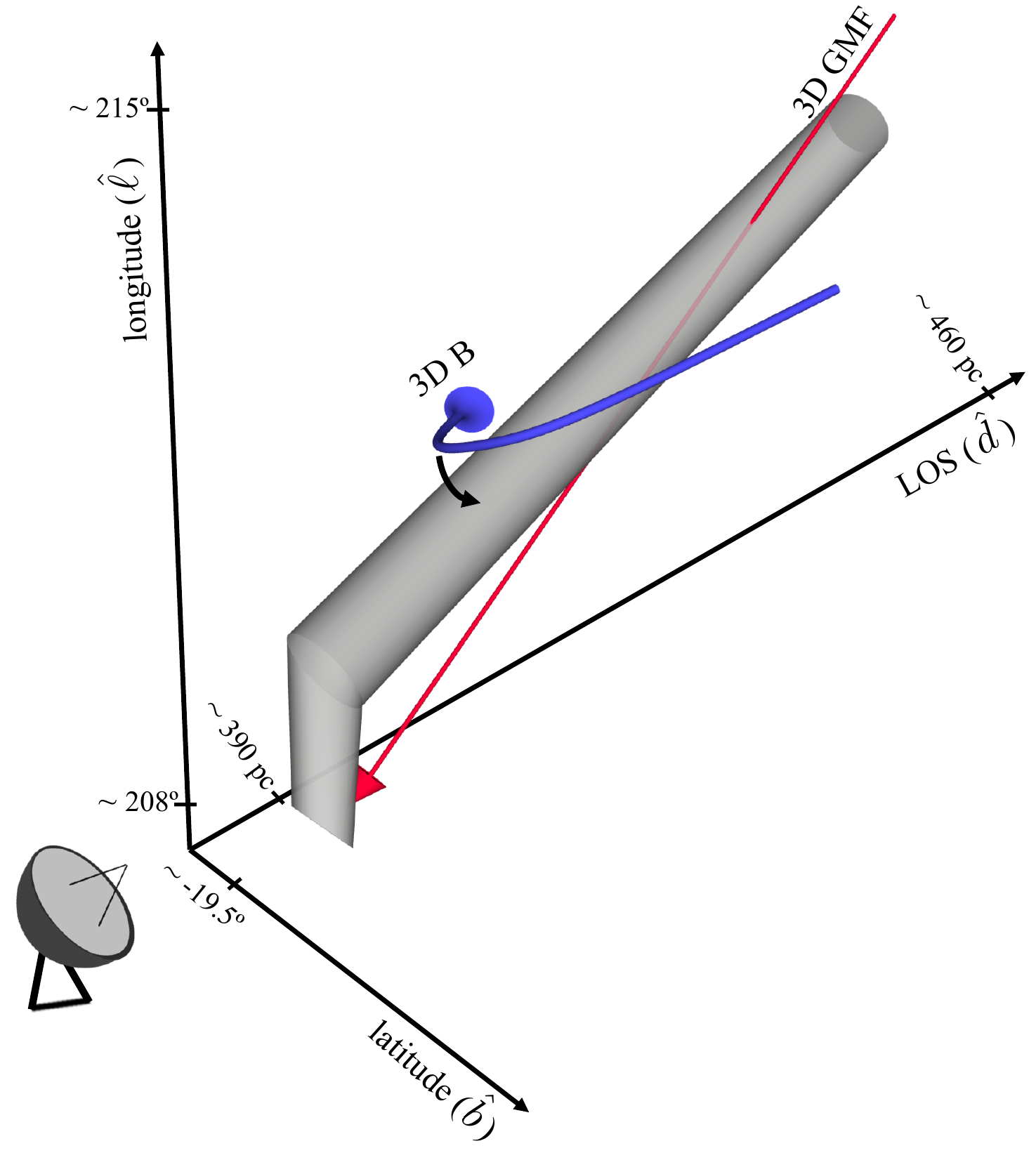}\\
\includegraphics[scale=0.45, trim={0cm 0cm 0.cm 0.cm},clip]{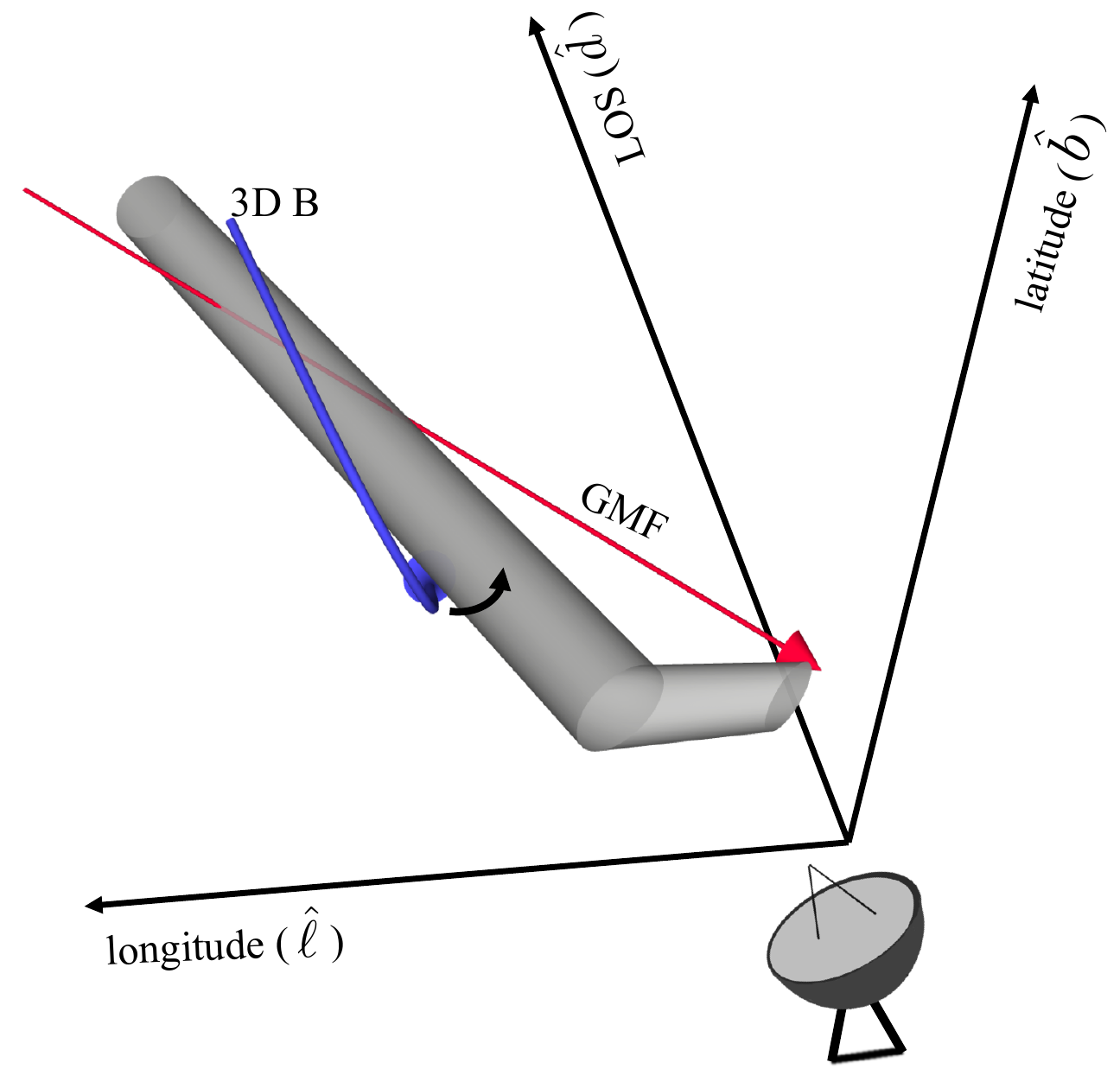}
\caption{Cartoon showing the 3D morphology of the Orion A magnetic field. The bent gray cylindrical shape and the red vector depict the Orion A filamentary structure and the mean 3D Galactic magnetic field in the region, respectively. 
The blue arrow shows the 3D magnetic field morphology of the cloud and is consistent with the \blos\ and \bperp\ observations shown in Fig.~\ref{fig:OrionBow}. \textbf{Top panel:} Projection of the cloud onto the plane of the sky. \textbf{Middle and lower panels:} Different viewing angles of the cloud. The black arrow denotes the direction in which rotations of up to maximum $50^{\circ}$ may be possible.} 
\label{fig:3DOrionField}
\end{figure}

To best describe the GMF vectors, we employ a frame of reference in this region (see Fig.~\ref{fig:3DOrionField}), with its axes pointing in the increasing directions of longitude ($\hat{\ell}$), latitude ($\hat{b}$), and distance ($\hat{d}$) at Orion A's plane-of-sky location (where $\hat{\ell}$, $\hat{b}$, and $\hat{d}$ are unit vectors). The GMF direction in this region can be described as a unit vector of  $-0.7 \hat{\ell} -0.1 \hat{b} -0.7 \hat{d}$. This vector appears mostly parallel to Orion A when projected onto the plane of the sky, but has a large component along the line of sight (pointing toward us). This vector is consistent with previous GMF studies~\citep[e.g.,][see their Fig.~6]{VanEcketal2011}, with the same inclination angle and direction~\citep[45$^{\circ}$ to 50$^{\circ}$ with respect to the plane of the sky; e.g.,][]{OrenWolfe1995, Heiles1997}. 

%------------------------------------------------------
%---------------------------------------------------------

We also need to account for the inclination angle of Orion A when reconstructing its 3D magnetic field. \citet{Grossschedletal2018} demonstrated that the majority of Orion A has a high inclination angle with respect to the plane of the sky. They divided the cloud into a head (longitudes less than $\sim 210^{\circ}$) and a tail (longitudes greater than $\sim 210^{\circ}$, as depicted in Fig.~\ref{fig:GMFOrionA}),  
with the head running approximately parallel to the plane of the sky and the  tail trending into the line of sight at an approximately $70^{\circ}$ inclination angle, as illustrated in Fig.~\ref{fig:3DOrionField}. 

We reconstruct an arc-shaped morphology as depicted in  Fig.~\ref{fig:3DOrionField}, by connecting the approximate GMF vectors to the \blos\ observations via an arc that accounts for the 3D orientation of the cloud and matches the \bperp\ morphology and the relative \blos\ strengths.  
The initial magnetic field (the GMF) is depicted in this figure as a red vector pointing toward us, and the filament is shown with a gray cylindrical shape. The blue vector illustrates the reconstructed arc-shaped magnetic field morphology, which looks semi-convex (from our point of view) and points in the direction of decreasing latitude.  

This 3D magnetic field morphology is an approximate  large-scale configuration that ignores smaller-scale variations and fluctuations, enabling us to determine the direction and orientation of the arc, as well as the direction of magnetic fields projected onto the plane of sky. Therefore, small variations of this field could be possible as long as they remain consistent with the \blos\ and \bperp\ observations, as well as the GMF vectors. While the majority of these variations are insignificant, rotations up to maximum $50^{\circ}$ along the black arrow shown in the middle and lower panels of Fig.~\ref{fig:3DOrionField} may be possible (also see Fig.~\ref{fig:3D2OrionField}), covering possible field shapes  from the front to the back side of the cloud (that are consistent with the observations). The field shapes in this possible range remain semi-convex from our point of view and point in the decreasing latitude direction when projected onto the plane of the sky. 

Moreover, while we did not directly consider field strengths  when reconstructing the field lines, the results are consistent with observed strengths; the cloud has a stronger line-of-sight field component to its Galactic north side than to its Galactic south side. This is consistent with the \blos\ observations of \citet{Tahanietal2018}  and Zeeman measurements of \citet{Heiles1997}. The error-weighted average strength of the \blos\ observations in both studies is twice as strong on the cloud's Galactic north side as on the cloud's Galactic south side.
Finally, while we believe that this 3D field is the most probable and natural field morphology for the Orion A cloud,
it does not completely rule out other possibilities. Future observations with a high rotation measure source density and improved \bperp\ resolution will allow us to more accurately and precisely determine the 3D fields.

%------------------------------------------------------
%---------------------------------------------------------

\section{Discussion}
\label{Sec:ArcFormation}

Our reconstructed 3D magnetic morphology is consistent with the SCI cloud-formation model, in which the surrounding environment (e.g., bubbles, events, structures, or sheet-like clouds) influenced and bent the field lines. In this section we discuss the formation of this arc-shaped morphology and environmental influences that shaped it.

\subsection{Bubbles influencing the Orion A cloud}
\label{Sec:bubbles}

\begin{figure}[htbp]
\centering
\includegraphics[scale=0.8, trim={0cm 0cm 0.5cm 0.5cm},clip]{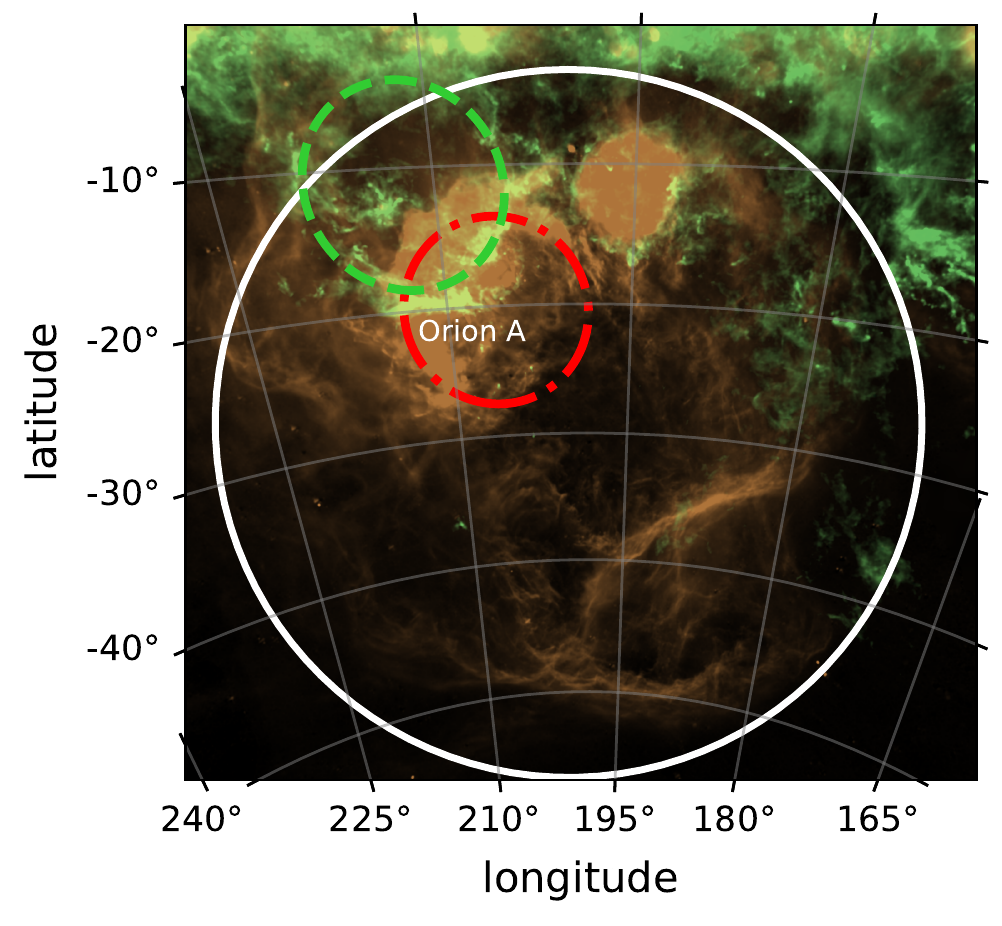}
\caption{Observations of the Orion-Eridanus superbubble and its nested bubbles. The white, green dashed, and red dash-dotted circles illustrate the approximate outline of the Orion-Eridanus superbubble, dust ring, and Barnard's loop on the plane of the sky, respectively. The light-green and orange background colors represent the observations of thermal dust (at 545\,GHz) obtained by the Planck Space Observatory and H$\alpha$ emission~\citep{Finkbeiner2003}, respectively. The bright green region shows Orion A. } 
\label{fig:HalphaDustBubbles}
\end{figure}

\begin{figure}[htbp]
\centering
\includegraphics[scale=0.15, trim={0cm 0cm 0cm 0cm},clip]{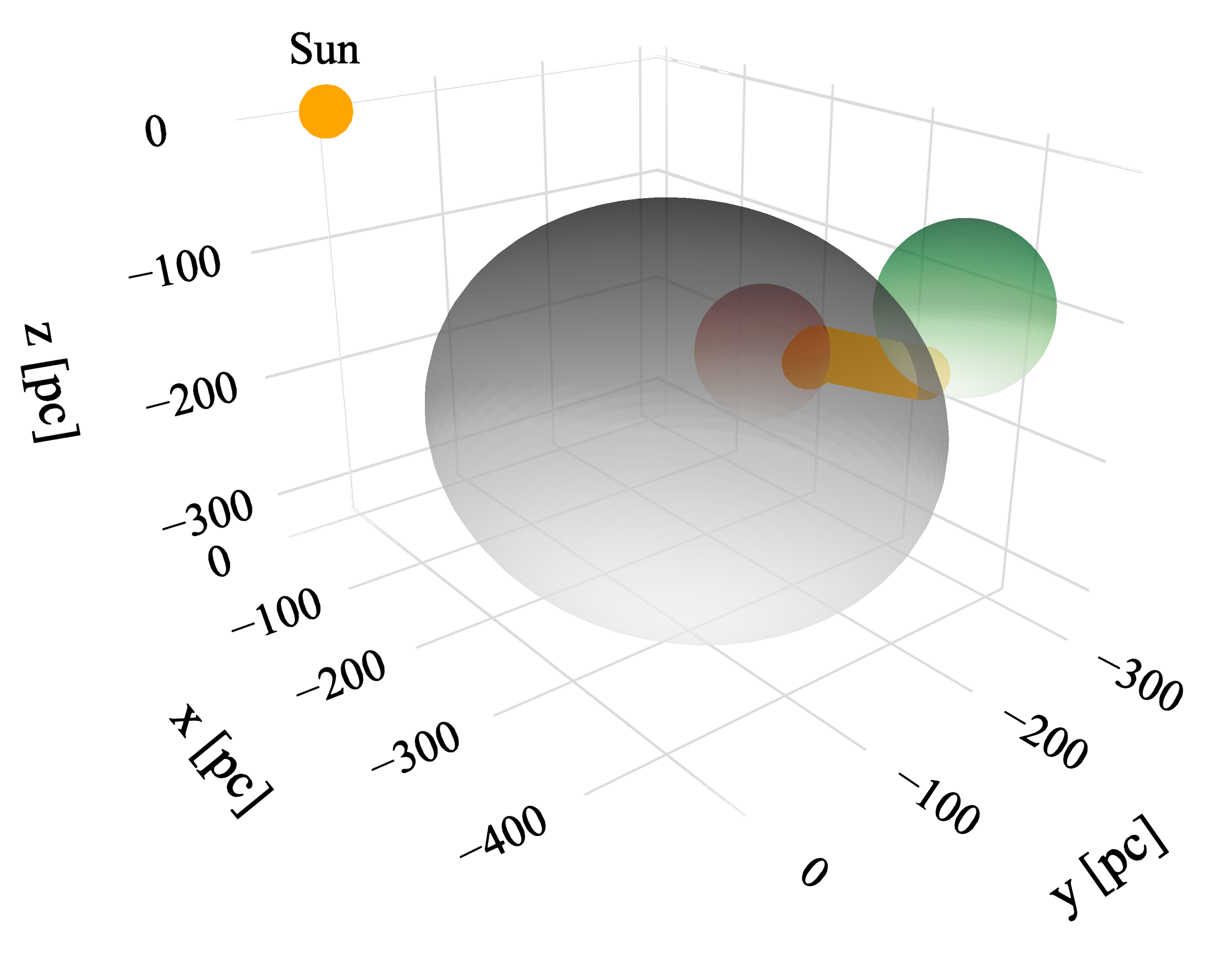}
\caption{A simplified 3D approximation of the Orion-Eridanus superbubble and its associated bubbles. Orion A is shown as an orange bent cylinder and the Sun is depicted as a yellow circle. The gray ellipsoid and red and green spheres represent the Orion-Eridanus superbubble, Barnard's loop, and the dust ring, respectively. Due to uncertainties in models of the Orion-Eridanus superbubble, the dust ring might also be embedded within the superbubble. 
} 
\label{fig:3DShells}
\end{figure}

\textbf{Orion-Eridanus superbubble:} Orion A \citep[located at $432 \pm 2$\,pc from the Sun;][]{Zuckeretal2019}  was likely shaped by energy release from the Orion OB association~\citep{Bally2008}. From the Sun, the cloud is seen in the projected interior of the Orion-Eridanus superbubble. Its distance puts it on the far side of the superbubble, but its location within, along, or beyond the bubble rim remains uncertain because of the complex composite structure of the superbubble \citep{Ochsendorfetal2015, Joubaudetal2019}. The stellar content~\citep{Vossetal2010}, stellar age gradient~\citep{BouyAlves2015, Zarietal2017}, and X-ray temperature gradient~\citep{Joubaudetal2019} all indicate that the bubble has likely evolved in time and space from a near (150-200\,pc away) to far distance (close to Orion A and B) after a 10- to 20-Myr-long series of supernova events. 
The approximate outline of the superbubble~\citep[based on][]{Ochsendorfetal2015} is depicted in Fig.~\ref{fig:HalphaDustBubbles} as a white circle, with the background light-green and orange colors representing thermal dust (at 545\,GHz) and H$\alpha$ observations, respectively. 
Figure~\ref{fig:3DShells} illustrates a rough 3D approximation of the location of this superbubble as a gray ellipsoid, based on the 3D models of \citet{Ponetal2014} and \citet{Ponetal2016}.

\citet{Grossschedletal2020}, \citet{Kounkel2020}, and \citet{Pellizaetal2005} found coherent stellar proper motion associated with the Orion region and the superbubble.  Coherent proper motions observed by \citet{Grosschedletal2020b} and \citet{Kounkel2020} are associated with the young stellar objects in the Orion A and B clouds, implying the influence of a feedback-driven event.  \citet{Grosschedletal2020b} refers to this event as Orion-BB (big blast), whereas \citet{Kounkel2020} links the coherent velocities to the Barnard's loop. In general, these coherent velocities are likely indicative of feedback-driven (e.g., supernovae) impacts on the Orion A and B clouds and their young stellar objects.

\textbf{Barnard's loop:} Barnard's loop~\citep{Barnard1894} is a complete bubble located within the Orion-Eridanus superbubble \citep{Ochsendorfetal2015}. The Barnard's loop bubble, with an estimated age of $3\times 10^5$\,yr, is expanding at a velocity of 100\,km\,s$^{-1}$~\citep{Ochsendorfetal2015}, while the Orion-Eridanus superbubble has an expansion velocity of $\sim 20$\,km\,s$^{-1}$~\citep{Joubaudetal2019}. 
This bubble is located between 340\,pc and 400\,pc from  us~\citep{Grossschedletal2020}. Using Barnard's loop and the Orion A cloud's distances, we find that the Barnard's loop bubble is likely in contact with (and has interacted with) Orion A's head but not its tail, as depicted in Fig.~\ref{fig:3DShells}. This could account for the tilt  of Orion A's head  and the fact that the star formation rate in Orion A's head is an order of magnitude greater than in its tail~\citep{Grosschedletal2020b}. We predict that as the Barnard's loop bubble expands, it will interact with Orion A and trigger a new star-formation sequence in the tail. 

\textbf{Orion dust ring:} \citet{Schlaflyetal2015} map the 3D dust and find a dust ring (indicating a bubble origin) between 400\,pc and 550\,pc (from us) in the Orion complex region. This dust ring is depicted in Fig.~\ref{fig:HalphaDustBubbles} as a green-dashed circle and is seen at more positive latitudes compared to Orion A.  \citet{Schlaflyetal2015} estimate the age of the bubble to be around 10 Myr or greater, because they find no evidence of H$\alpha$ associated with the ring. A closer examination of Fig.~\ref{fig:HalphaDustBubbles}, however, indicates that this ring may be visible in  H$\alpha$ observations. We roughly approximate the ring (and its original bubble) as a sphere (depicted in green in Fig.~\ref{fig:3DShells}) with a radius of 150\,pc and a center location of $\ell = 212^{\circ}$, $b = -11.5^{\circ}$, and $d = 175$\,pc.

%------------------------------------------------------
%---------------------------------------------------------

 % %-----------------------------------------------
% %----------------------------------------------

\subsection{Gradual formation of the arc-shaped field morphology}
\label{Sec:gradual}

The bubbles detailed in Sect.~\ref{Sec:bubbles} have influenced the evolution (and the formation) of the Orion A cloud and its magnetic field lines. These influences can be summarized in two steps: 1) recurrent interactions with old events or bubbles  
pushed the Orion A cloud, its ambient environment, and the field lines (toward more positive latitudes), resulting in large-scale bending of the initial Galactic magnetic field, as predicted by the simulations of~\citet{KimOstriker2015}. 2) Subsequently, further interactions (likely by the dust ring) pushed the \HI\ gas around the Orion A cloud (toward more negative latitudes), creating a sharp arc-shaped magnetic morphology associated with the Orion A cloud, as depicted in Fig.~\ref{fig:3DOrionField}.

This two-step process is similar to that described in \citet{Tahanietal2022} for the Perseus molecular cloud (see Fig.~\ref{fig:SCIModifiedModel} and Appendix~\ref{sec:SCIModifiedSec}). In the first step, the Galactic magnetic fields are bent approximately tangential to expanding objects or bubbles (e.g., Orion-BB event or Orion-Eridanus superbubble). This results in magnetic field lines that are bent at Orion A's location, point toward us on its Galactic north side, and are mostly parallel to the plane of the sky on its Galactic south side (and perpendicular to the formed filament main axis). Subsequently, the interaction with the dust ring bends these field lines even more, causing them to point away from us on the cloud's Galactic south side.  
Therefore, the field lines on the Galactic north side of the cloud have a greater component along the line of sight compared to the Galactic south side of the cloud, as discussed in Sect.~\ref{Sec:reconstructed3DField}. 

To examine this view further, we estimate the gas and magnetic field pressures (see Appendix~\ref{Sec:quantitatives}) before the formation of the Orion A cloud and in its present state.  During the initial stages of the Orion A cloud formation, we find that the gas pressure is greater than the magnetic field pressure \citep[using a Galactic magnetic field strength of $\sim 5\,\mu$G, particle volume density of a few to 100\,cm$^{-3}$, and ambient temperature of 1000\,K and higher;][]{Joubaudetal2019}, implying  that recurrent supernovae can easily bend the magnetic field lines, as seen in simulations of \citet{KimOstriker2015}. Furthermore, using an average non-thermal velocity dispersion of a few ($\sim 3$)\,km\,s$^{-1}$, we find an Alfv\'en Mach number of maximum $\sim 3$, indicating that the field lines can retain a memory of the large-scale initial field morphology~\citep[with small-scale variations; cf.][]{HanZhang2007}. As a result, when the field lines interact with bubbles and events in this region (such as the Orion-Eridanus superbubble or the Orion-BB event) and bend, they remain largely coherent~\citep{KimOstriker2015}, rather than becoming completely distorted and perturbed~\citep{LiKlein2019}.  

To estimate the gas and magnetic field pressures in the Orion A molecular cloud in its current state, we use the error-weighted mean magnetic fields on two sides of the cloud (for detections pointing toward us and away from us), which are 87\,$\mu$G and $-45$\,$\mu$G, respectively, based on the observations of \citet{Tahanietal2018}. Using particle volume density and temperature values of $\sim 10^4$\,cm$^{-3}$~\citep[e.g.,][]{Castetsetal1990, Dutreyetal1993, JohnstoneBally1999a, JohnstoneBally1999b} and 25\,K~\citep[e.g.,][]{Mitchelletal2001, JohnstoneBally2006, Buckleetal2012, Schneeetal2014}, we find that the magnetic field pressure of the Orion A cloud  
is approximately one order of magnitude greater than its gas pressure  
(on both sides of the cloud). Using an average non-thermal velocity dispersion of  2\,km\,s$^{-1}$~\citep{Goicoecheaetal2020}, we find Alfv\'en Mach numbers of  $0.4$ and $0.8$ for the two sides of the Orion A molecular cloud. This implies that the field lines do not deform and if their magnetic field morphology changes (due to interaction with the environment), they retain a memory of their previous field morphology. 

This gradual bending of the field lines is consistent with the cloud-formation model of \citet{Inutsukaetal2015}. We also determine the \HI\ and CO line-of-sight velocities in this region (See Appendix~\ref{Sec:Velocities}) and find no significant offset between the two, indicating that the offset velocities caused by bubble interactions have dissipated or are mostly in the plane of the sky. Given the morphology of the Orion A cloud (which is inclined along the line of sight) and the fact that the GMF vectors are parallel to the longitude axis and point toward us along the line of sight, the interactions necessary for the formation of the arc-shaped magnetic morphology are mostly parallel to  
the plane of the sky. Therefore, we expect that any velocity offsets that may hint at arc-shaped magnetic field morphology formation (and perhaps are not yet completely dissipated) should be parallel to the plane of the sky. 

\section{Summary and conclusions}
\label{Sec:summary}

We determine the large-scale (and approximate) 3D magnetic field shape of the Orion A molecular cloud using Galactic magnetic field models, as well as present line-of-sight and plane-of-sky magnetic field observations.  
From our perspective, this 3D field is generally semi-convex and points toward the decreasing longitude and decreasing latitude directions in the plane of the sky.  
To our knowledge, this is the first time that the complete 3D large-scale ($\sim$ a few to $\sim 100$\,pc) magnetic field of the Orion A cloud has been reconstructed (including its plane-of-sky direction). We suggest that the Orion-Eridanus superbubble (or events within it, such as the Orion-BB event) and the dust ring in this region are largely responsible for the development of the arc-shaped magnetic field morphology of Orion A. 

Reconstructing the 3D magnetic field morphology of Orion A, which is consistent with the Planck observations when projected onto the plane of the sky, relies mainly on the initial Galactic magnetic fields and the \blos\ observations. This 3D magnetic field is a large-scale approximation that neglects smaller-scale distortions or entanglements in the field lines. Because we are interested in relatively large-scale and approximate fields, the presence of low-density regions in the foreground of Orion A~\citep[especially those smaller than the size of the Orion A cloud and the scales at which the \blos\ reversal was  observed;][]{Rezaeietal2020}  has no effect on our results. 

\begin{acknowledgements}
We would like to express our appreciation to the anonymous referee for their thoughtful comments that helped us improve the results and paper. MT is grateful for the helpful discussion with Pak-Shing Li. We have used \LaTeX, Python and its associated libraries including astropy~\citep{astropy} and plotly~\citep{plotly}, PyCharm, Jupyter notebook, SAO Image DS9, the Starlink~\citep{Starlink} software, the Hammurabi code, Shapr3D, and Adobe Draw for this work. For our line integration convolution plot, we used a Python function originally written by Susan Clark. QuillBot\footnote{\url{https://quillbot.com/}} was used for language editing purposes.  
The Dunlap Institute is funded through an endowment established by the David Dunlap family and the University of Toronto. J.L.W. acknowledges the support of the Natural Sciences and Engineering Research Council of Canada (NSERC) through grant RGPIN-2015-05948, and of the Canada Research Chairs program. Y.D. acknowledges the support of JSPS KAKENHI grant 18H01250. M.H. acknowledges funding from the European Research Council (ERC) under the European Union's Horizon 2020 research and innovation programme (grant agreement No 772663).

\end{acknowledgements}

\bibliographystyle{aa} 
\bibliography{biblio}

\appendix
\section{Data used in the study}
\label{Sec:Appdata}
In this study, we use estimates of the initial magnetic field direction,  line-of-sight magnetic fields, and velocities associated with each cloud. We use the \citet[][]{JanssonandFarrar2012} GMF model to determine the Galactic magnetic field.  We employ the catalog of \cite{Tahanietal2018} for the \blos\ magnetic field information, and available CO and \HI\ observations for the velocities.

\subsection{Galactic magnetic field}
 The \citet[][]{JanssonandFarrar2012} model includes a two-dimensional (2D) thin-disk field component that is tightly coupled to the Galactic spiral arms, an azimuthal/toroidal halo field component, and an X-shaped vertical/out-of-plane field component.    
To estimate the GMF, we use the Hammurabi program\footnote{\url{http://sourceforge.net/projects/hammurabicode/}}~\citep{Waelkensetal2009Hammurabi}, which  is a synchrotron modelling code, that has been used in different studies~\citep[e.g.,][]{Adametal2016, JanssonandFarrar2012}.  
We find the GMF vectors within a box around the Orion A cloud ($160\rm{\,pc}\times 160\rm{\,pc}\times 200\rm{\,pc}$), in the longitude range of $205^{\circ}$ to $218^{\circ}$ and latitude range of $-26^{\circ}$ to  $-13^{\circ}$. We set a resolution of one GMF vector per $2\,\rm{pc\,}\times 2\,\rm{pc\,}\times 2\,\rm{pc} $. 

\subsection{Line-of-sight magnetic field}

We employ the \blos\ observations of \citet[][]{Tahanietal2018}, in which they used Faraday rotation measurements to determine \blos\ of molecular clouds. They used RM point sources along with an on-off approach based on relative measurements to decouple the molecular clouds' contribution to RM from that caused by the rest of the Galaxy.  
They then calculated the strength of \blos\ employing a chemical evolution code and the \citet{Kainulainenetal2009} extinction maps~\citep[for more details, see][]{Tahanietal2018}. 

\subsection{Velocity information}

We consider both the available CO and \HI\ velocities to explore the line-of-sight velocities of the cloud and its surrounding environment.  
We obtain \HI\ velocity information  from the all-sky database of the \HI\ $4\pi$ Survey \citep[HI4PI;][]{HI4PICollaboration2016}, which is based on the Effelsberg-Bonn \HI\ Survey \citep[EBHIS;][]{EBHIS_Kerpetal2011, EBHIS_Windeletal2016} and the Galactic All-Sky Survey \citep[GASS;][]{McClureGriffithsetal2009}.
We use the radial velocities from the \citet{Dameetal2001} carbon monoxide survey, to determine the cloud's CO velocity. This catalog is a survey of the $^{12}$CO J(1-0) spectral line of the Galaxy.  

\section{3D magnetic field sensitivity}
\label{Sec:3DFieldSens}

\begin{figure}[htbp]
\centering
\includegraphics[scale=0.45, trim={0cm 0cm 0.cm 0.5cm},clip]{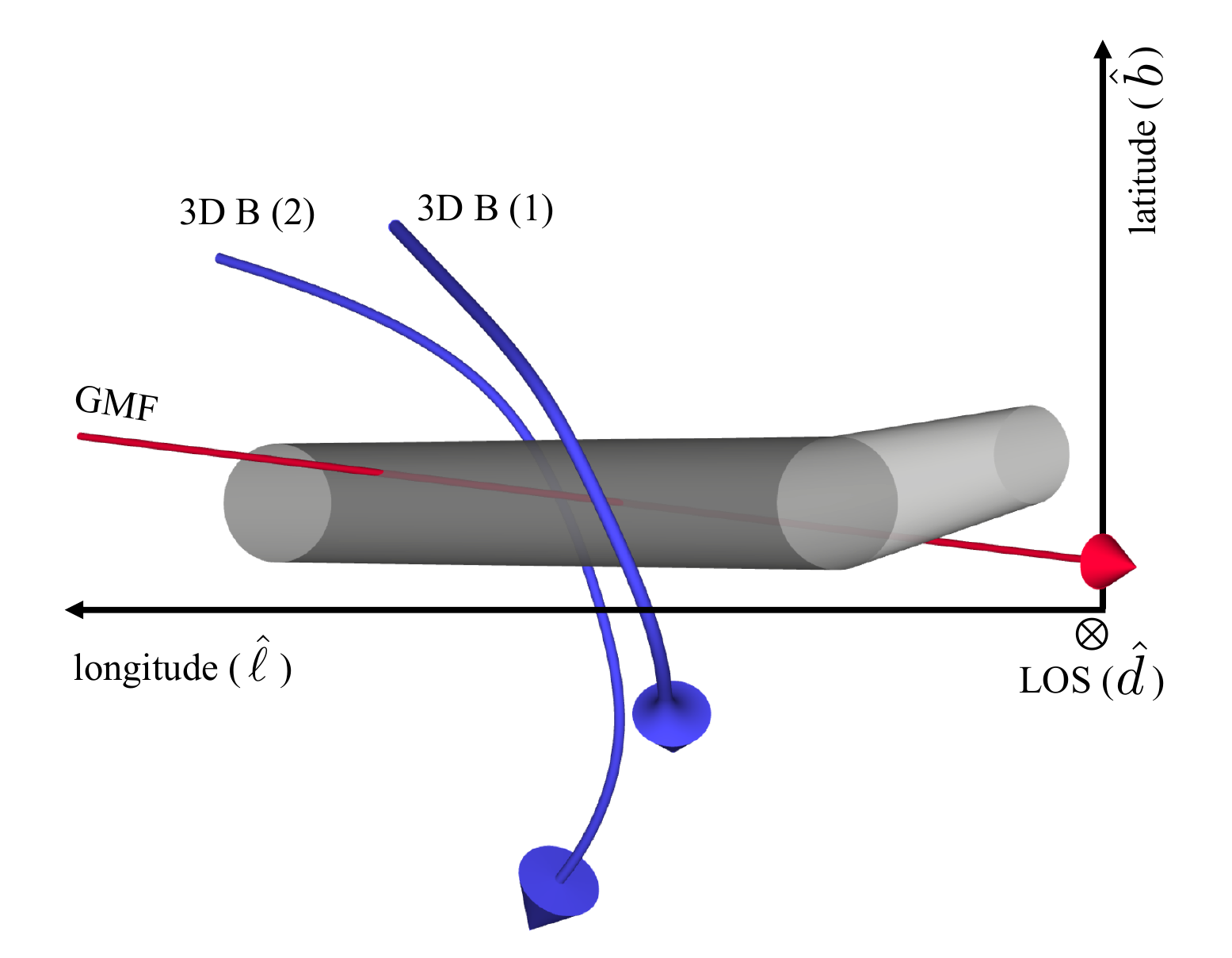}\\
\includegraphics[scale=0.49, trim={0cm 0cm 0.cm 0cm},clip]{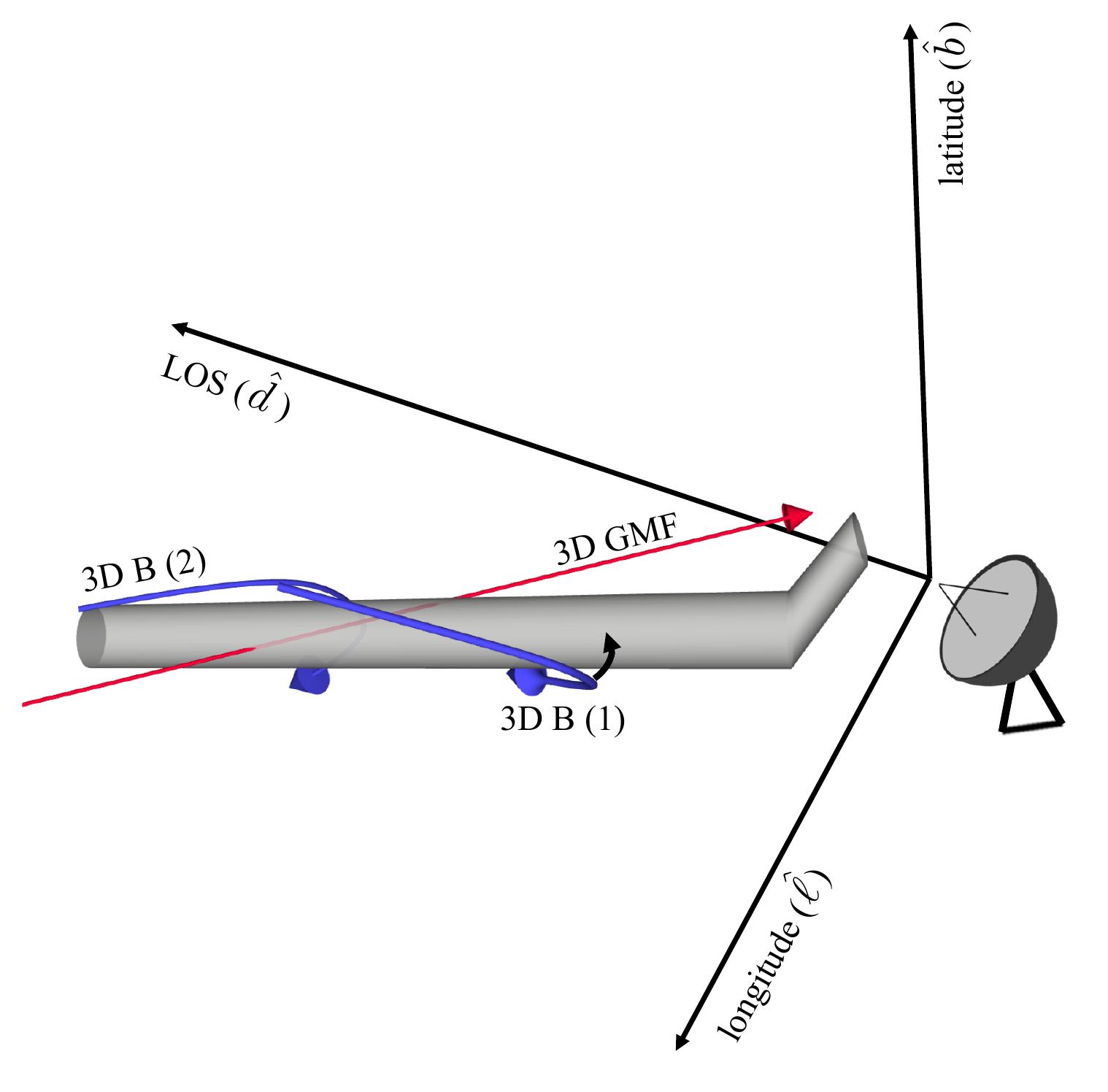}\\
\includegraphics[scale=0.45, trim={0cm 0cm 0.cm 0cm},clip]{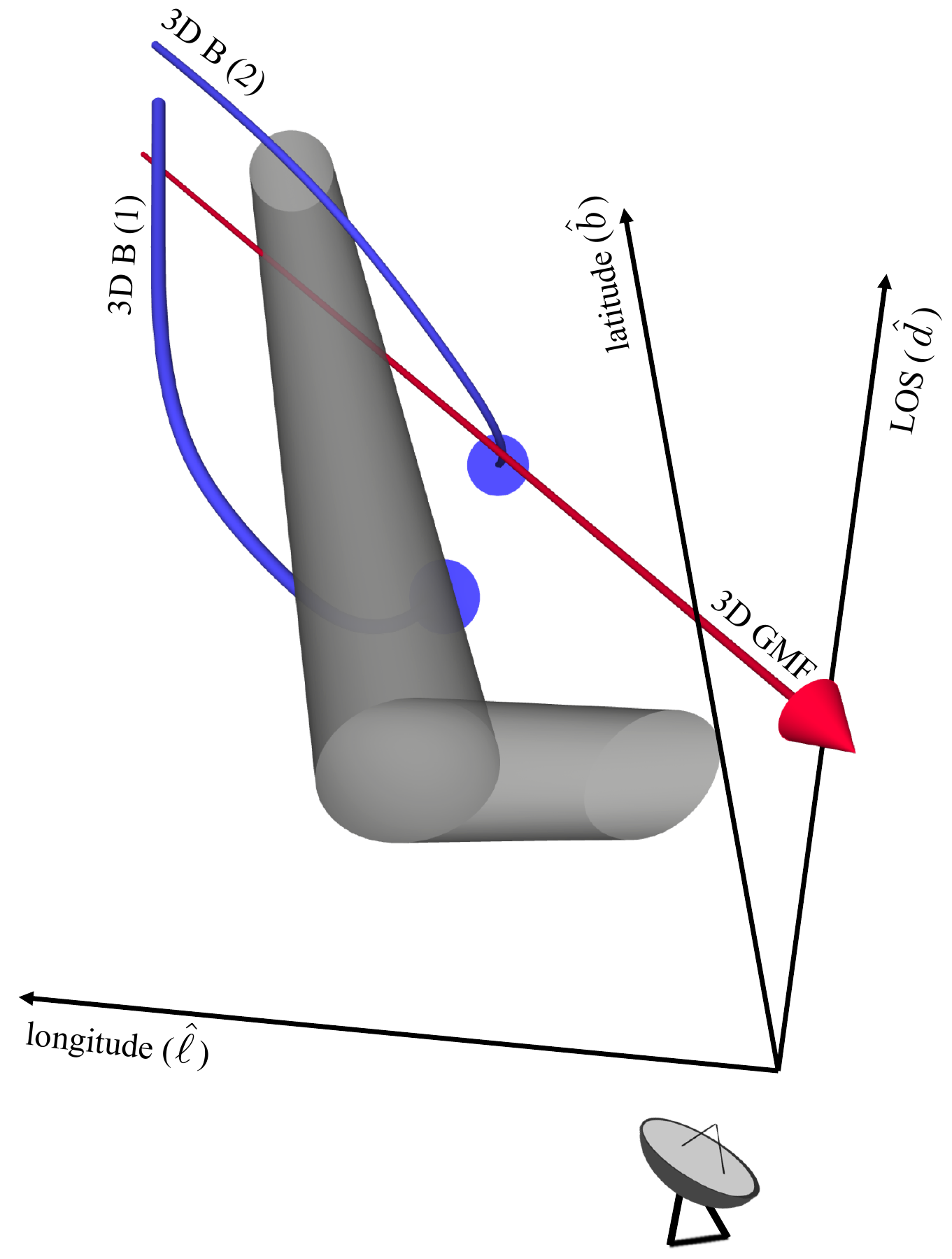}
\caption{Cartoon showing the sensitivity of the reconstructed 3D magnetic field. The Galactic magnetic field, the Orion A cloud, and the reconstructed field are shown with the red vector, bent gray cylindrical shape, and the blue 3D B (1) arrow, respectively. \textbf{Top panel:} Projection onto the plane of the sky. \textbf{Middle and lower panels:} Different viewing angles of the cloud. The black arrow in the middle panel illustrates the direction of rotation to obtain 3D B(2), which remains consistent with the observations and is semi-convex from our point of view.
} 
\label{fig:3D2OrionField}
\end{figure}

As discussed in Sect.~\ref{Sec:reconstructed3DField}, the reconstructed 3D field is an approximation of the large-scale field, which allows for small-scale variations such as small rotations along the black arrow shown in the middle panel of Fig.~\ref{fig:3D2OrionField}. Rotations in other directions, on the other hand, are more sensitive and limited, as they do not produce the observed \blos\ or \bperp . We emphasize that these variations must remain consistent with the GMF, as well as the \blos\ and \bperp\ observations. The magnetic field that we reconstructed\footnote{The .obj files are available at \url{https://github.com/MehrnooshTahani/OrionA3DMagneticFields}}  in Sect.~\ref{Sec:reconstructed3DField} is depicted as 3D B (1) in Fig.~\ref{fig:3D2OrionField}. Examining a large number of 3D field varieties, we find that rotations along the black arrow up to maximum $ 50^{\circ}$ may be possible (covering possible field morphologies from the front to the back side of the cloud), resulting in the arc-shaped field depicted as 3D B(2) in Fig.~\ref{fig:3D2OrionField}, which is semi-convex from our perspective and points in the direction of decreasing latitude.

%------------------------------------------------------
%---------------------------------------------------------

\section{Velocity observations}
\label{Sec:Velocities}

To investigate the line-of-sight velocities of the cloud thoroughly, we first calculate the line-of-sight Galactic rotation velocities.  
We find a Galactic rotation velocity (LSR) of $\sim  +4$\,\kms\ for the Orion A cloud ($l \simeq 211^{\circ}$ and $d \simeq 432$\,pc), using the model of~\cite{Clemens1985_GalRot} with the IAU standard values of the solar distance from the Galactic center 
(8.5\,kpc) and orbital velocity (220\,\kms ).

For \HI\ velocities, we select locations that have a single \HI\ peak emission and exclude those with multiple peaks, absorption, or self-absorption, as discussed in \citet{Tahanietal2022}.  
Similarly, we pick points whose CO spectrum can be described primarily by a single Gaussian fit. For more accuracy, we take \HI\ and CO data from similar coordinates. These \HI\ and CO velocities agree with other studies at similar coordinates~\cite[e.g.,][]{Kongetal2015, Riceetal2016, Ishiietal2019, Maetal2019}. We note that average CO and HI velocities along the main axis of the cloud are close to Galactic rotation velocities at this location ($\sim 4$ to $6$\,\kms ).

Subsequently, we can  find the molecular cloud velocities (CO) in the co-moving frame of the \HI\ gas using:
\begin{equation}
v_{\text{CO - H\protect\scaleto{$I$}{1.2ex}}} = v_{\text{CO, LSR}} - v_{\text{H\protect\scaleto{$I$}{1.2ex}, LSR}},
\end{equation} 
where \vcoHi\ is the cloud CO velocity in the co-moving \HI\ frame, \vco\ is the CO velocity in the local standard of rest (LSR) frame and \vhi\ is the \HI\ velocity of the region in the LSR frame. To account for the velocity gradients and fluctuations along the cloud, we find \vcoHi\  on the cloud point by point (considering the peak emission of CO and \HI ) and then take an average.  This obtained value is $0.0 \pm 0.5$\,\kms\ (where the uncertainty value is the standard deviation of the \vcoHi\ points), indicating that within the uncertainty range of the observations, there is no significant line-of-sight offset between the \HI\ and CO velocities in the Orion A cloud \cite[also consistent with Fig.~1 of][]{ImaraBlitz2011}. 

This zero~\kms\ line-of-sight average CO velocity in the co-moving \HI\ frame emphasizes the importance of considering the plane-of-sky environmental influences and their effect on the GMF vectors.  Furthermore, since the GMF vectors fall parallel to the cloud on the plane of the sky and the cloud has a large inclination angle, it is even more critical to consider plane-of-sky influences and events \citep[e.g., coherent velocities observed by][]{Grossschedletal2020} in order to fully understand the evolution of 3D magnetic fields in this region.

\section{Orion bubbles}
\label{sec:SCIModifiedSec}
\begin{figure*}[htbp]
\centering
\includegraphics[scale=0.6, trim={0cm 0cm 0cm 0cm},clip]{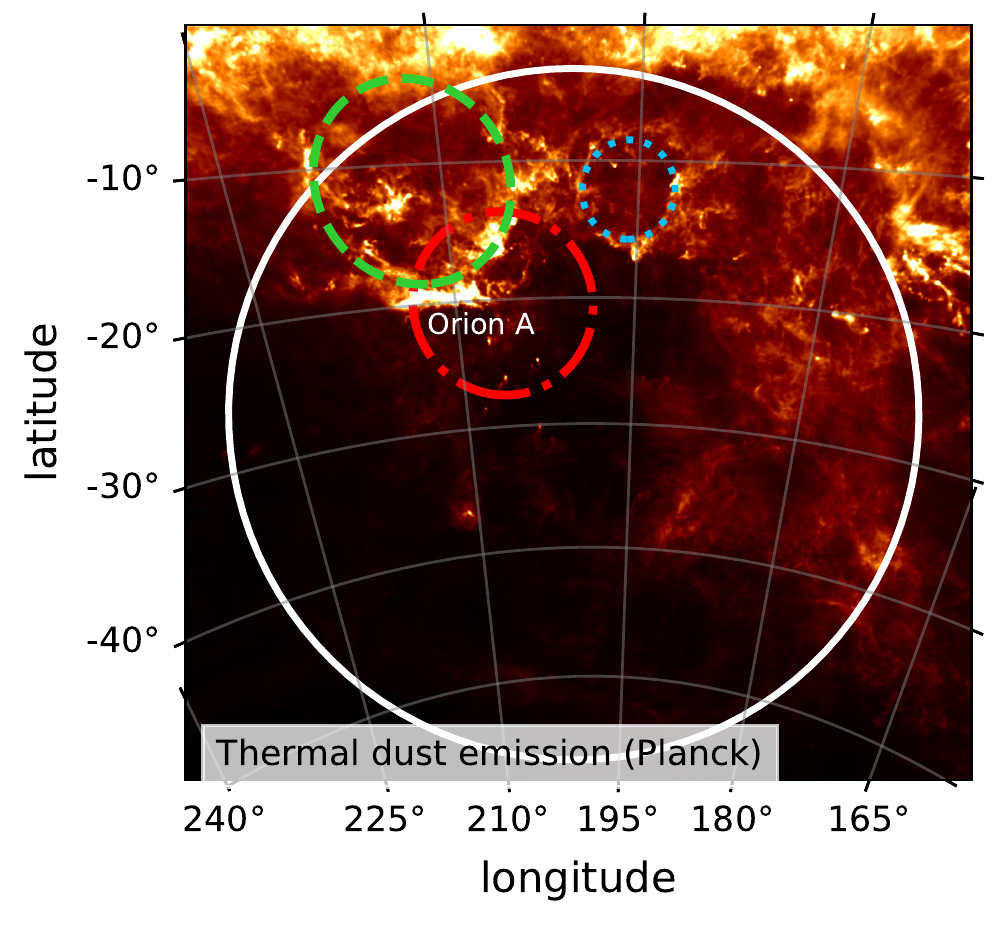}
\includegraphics[scale=0.6, trim={0cm 0cm 0cm 0cm},clip]{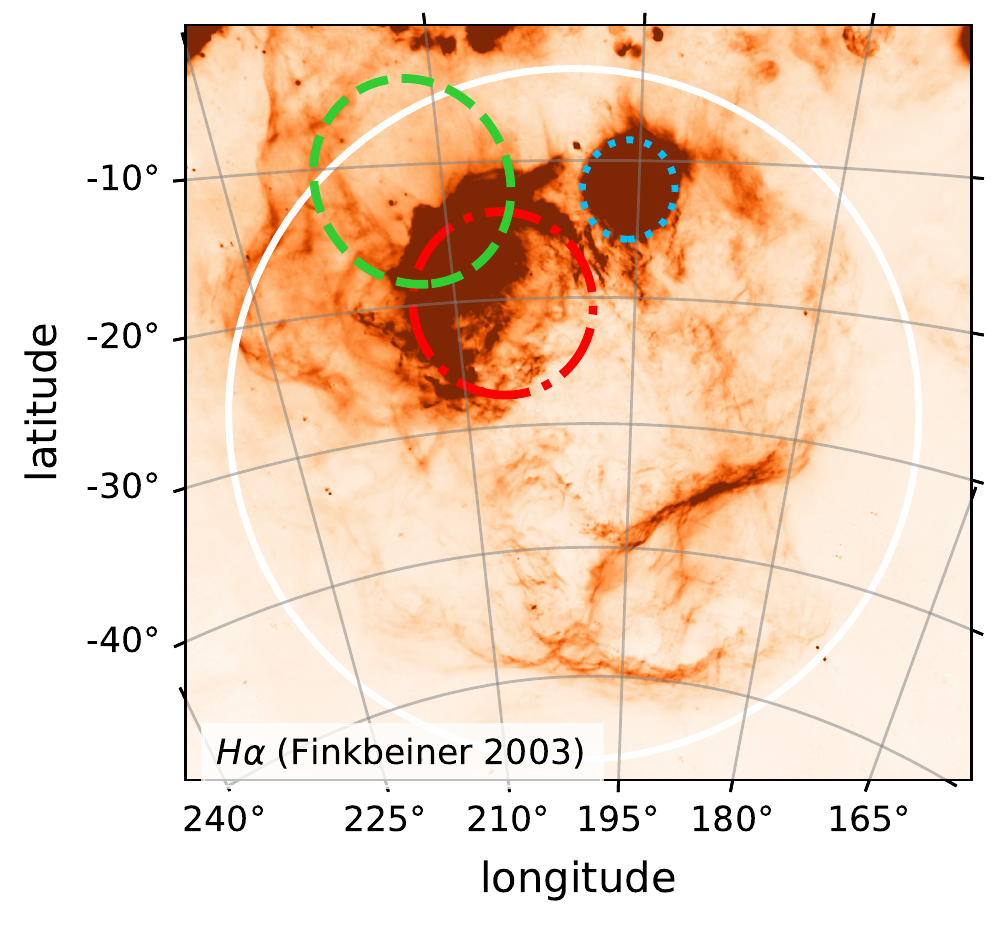}
\includegraphics[scale=0.6, trim={0cm 0cm 0cm 0cm},clip]{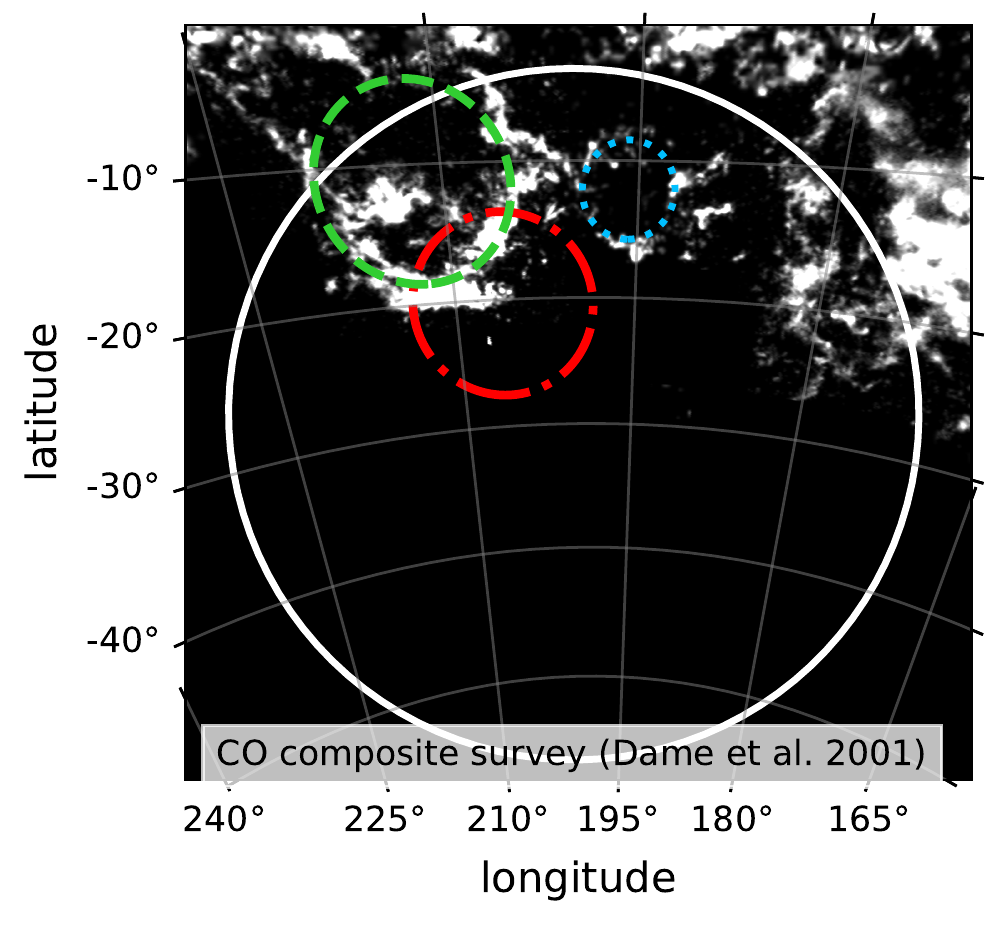}\\
\includegraphics[scale=0.6, trim={0cm 0cm 0cm 0cm},clip]{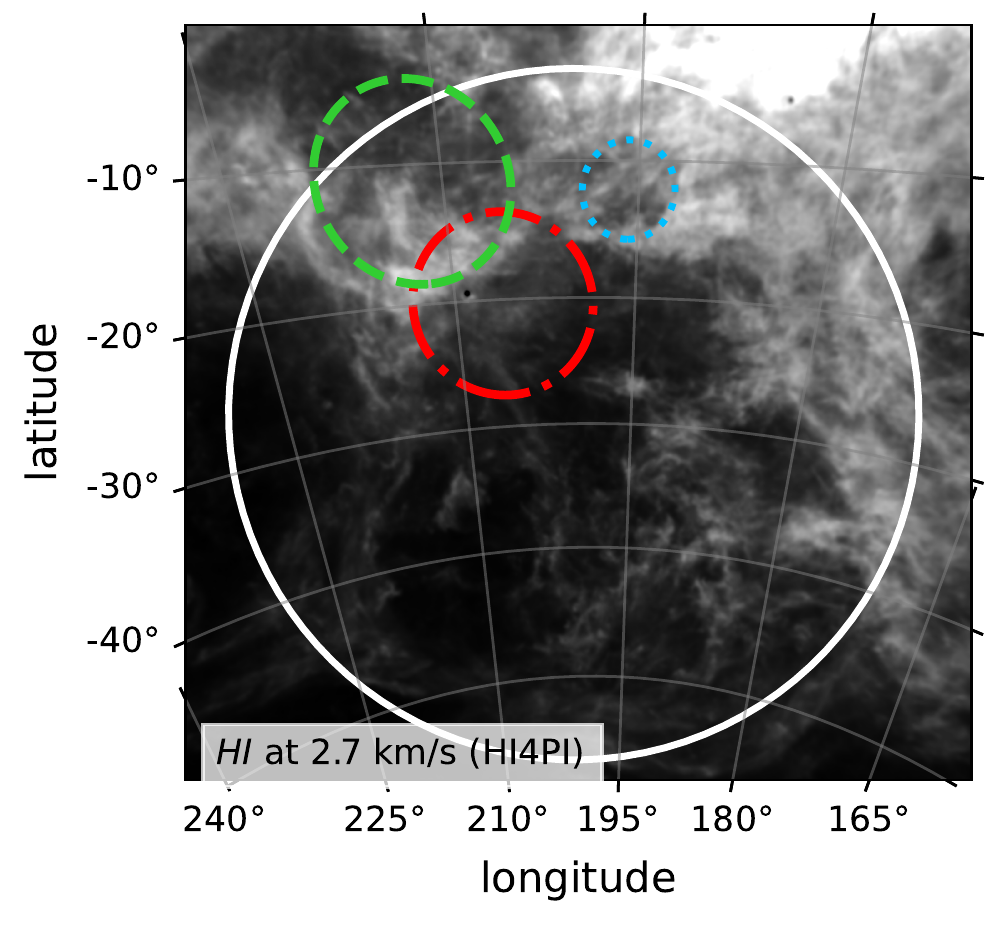}
\includegraphics[scale=0.6, trim={0cm 0cm 0cm 0cm},clip]{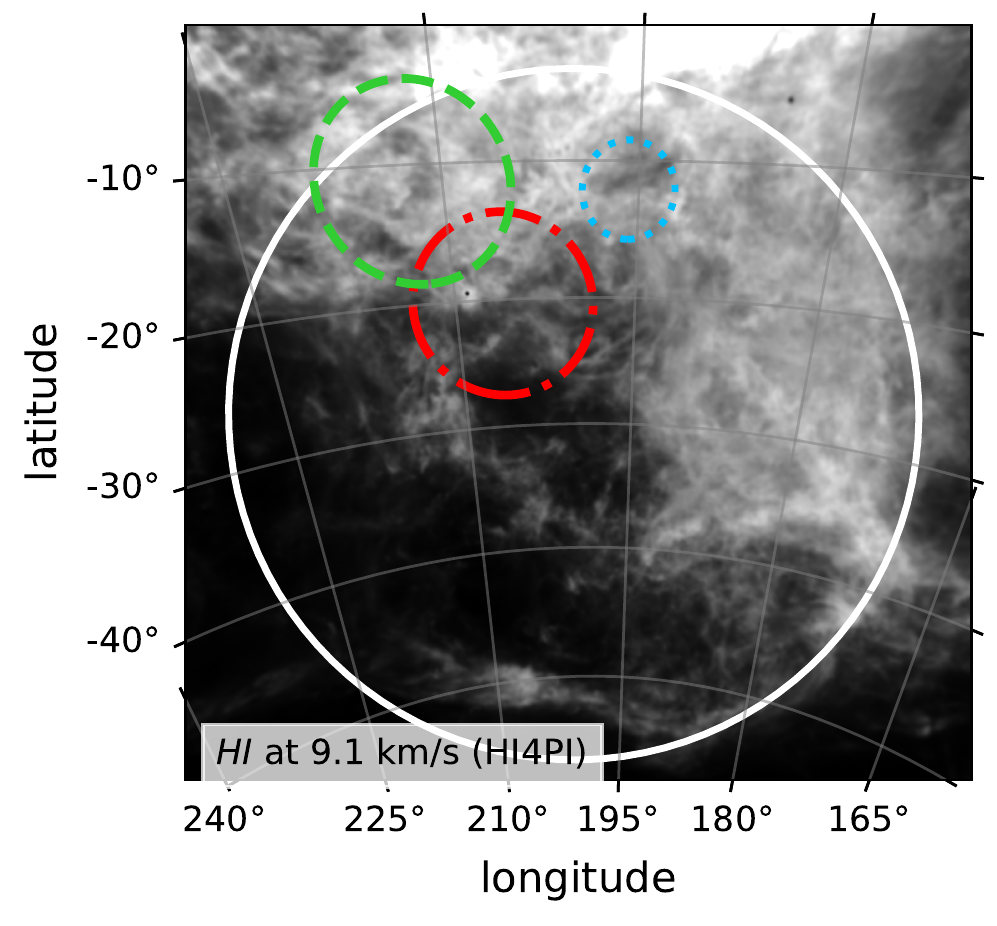}\\
\caption{Multi-wavelength observations of the Orion-Eridanus superbubble. The Orion-Eridanus superbubble, dust ring, Barnard's loop, and the $\lambda$ Orionis ring are depicted as white, green dashed, red dash-dotted, and blue dotted circles, respectively.  
\textbf{Top-left panel:} Thermal dust observations from the Planck Space Observatory. 
\textbf{Top-middle panel:} $H\alpha$ observations \citep{Finkbeiner2003}. \textbf{Top-right panel:}  Composite CO survey of \cite{Dameetal2001}. \textbf{Lower panel:} \HI\ observations at  velocities of  $2.7$\,km\,s$^{-1}$ and $9.1$\,km\,s$^{-1}$ from HI4PI.  } 
\label{fig:MultiWaveAllBubble}
\end{figure*}

We discussed the bubbles that influence the Orion A cloud in Sect.~\ref{Sec:bubbles} and Sect.~\ref{Sec:gradual}. The presence of these bubbles is further demonstrated by multi-wavelength observations shown in Fig.~\ref{fig:MultiWaveAllBubble}.  Thermal dust, H$\alpha$, CO, and \HI\  observations are illustrated in this multi-panel figure, with the Orion-Eridanus superbubble, dust ring, Barnard's loop, and the $\lambda$ Orionis ring outlined as white, green dashed, red dash-dotted, and blue dotted circles, respectively. 

\begin{figure*}[htbp]
\centering
\includegraphics[scale=0.25, trim={0cm 0cm 0cm 0cm},clip]{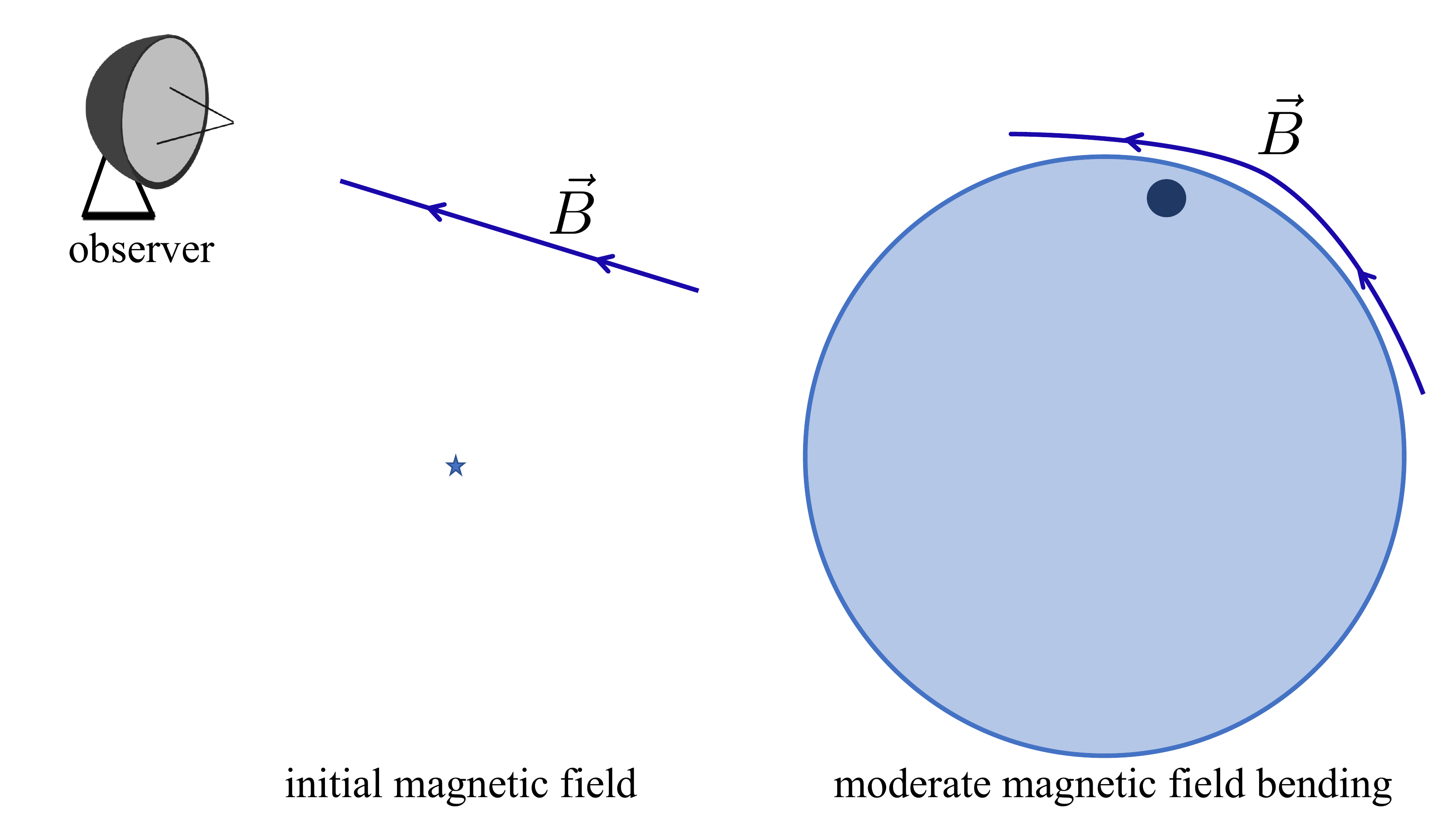}
\hspace{1cm}
\includegraphics[scale=0.25, trim={0cm 0cm 0cm 0cm},clip]{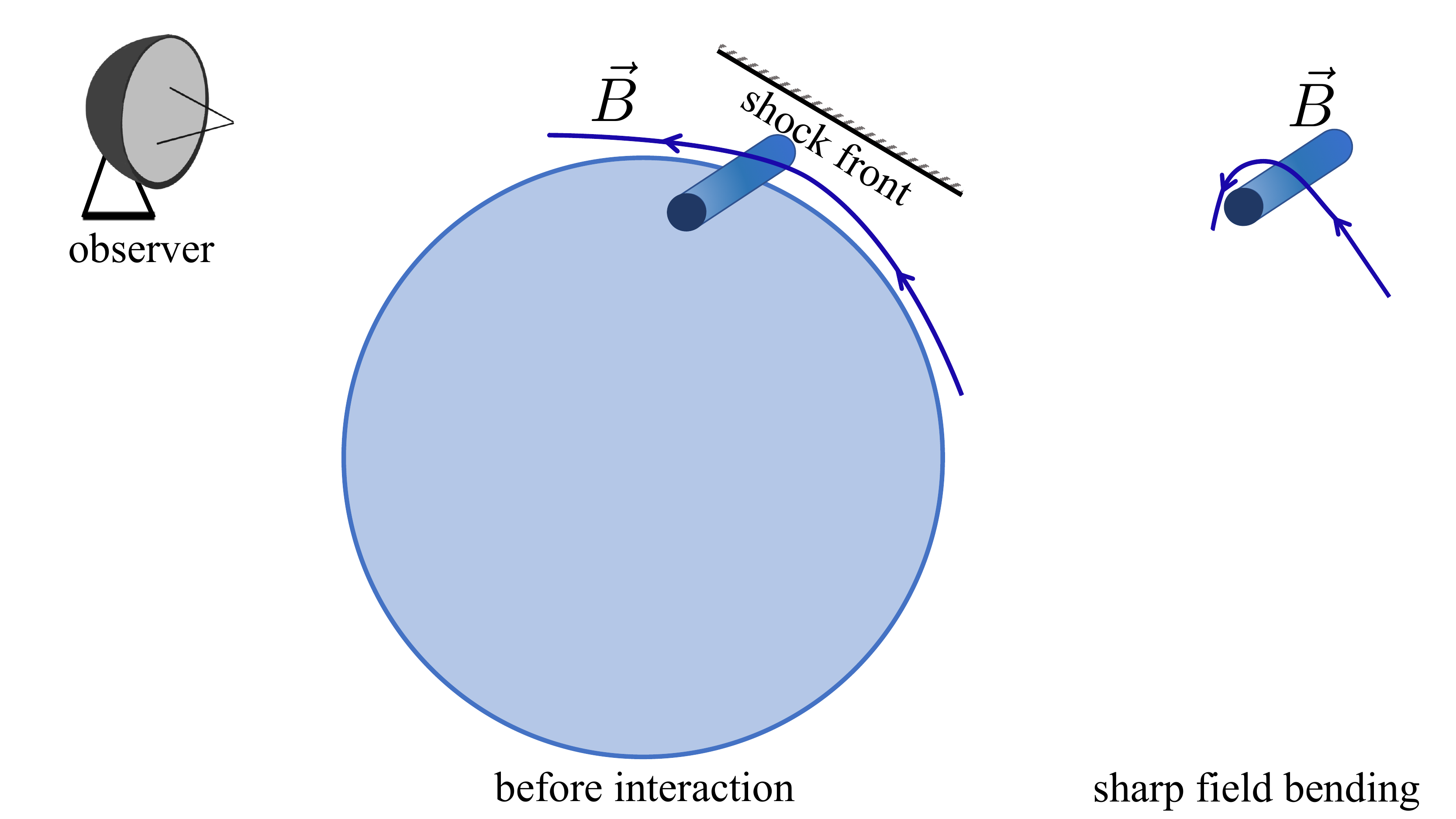}
\caption{Formation of the arc-shaped magnetic field morphology in the Orion A cloud. The large light blue circle represents a bubble within the Orion-Eridanus superbubble, or the Orion-Eridanus superbubble itself (progenitor(s) represented by the star sign). 
The small dark blue cylinder illustrates the Orion A filamentary molecular cloud. The 3D magnetic configurations cannot be accurately represented in 2D and a more accurate representation of the field lines is shown in  Fig.~\ref{fig:3DOrionField}. \textbf{Left panel}: The straight \vec{B} represents the direction of the initial magnetic field prior to interaction with the bubble/event, while the large-scale mild bending of the magnetic field morphology after interaction with the bubble/event is represented by the curved \vec{B}. \textbf{Right panel}: Formation of the arc-shaped magnetic field morphology around the Orion A molecular cloud due to further interactions, possibly with the dust ring.
}
\label{fig:SCIModifiedModel}
\end{figure*}

In Sect.~\ref{Sec:gradual}, we discussed the gradual evolution of initial magnetic field lines, resulting in the arc-shaped morphology depicted in Fig.~\ref{fig:3DOrionField}. This evolution process is summarized in  Fig.~\ref{fig:SCIModifiedModel}: 1) The first step (left panel) involves the bending of Galactic magnetic fields toward positive latitudes (by the Orion-Eridanus superbubble, or bubbles and events within, such as the Orion-BB event). At this stage field lines are bent at Orion A's location, pointing toward us on its Galactic north side, and mostly parallel to the plane of the sky on the Galactic south side of the cloud (on the plane of the sky appearing mostly perpendicular to the main axis of the filament that is formed). 
2) Following that (right panel), additional environmental influences, such as the dust ring, interact with the cloud's surrounding \HI\ gas, further bending these field lines, causing them to point away from us on the cloud's Galactic south side. 

 % %-----------------------------------------------
 % %-----------------------------------------------

%----------------------------------------------------
%----------------------------------------------------
 
 \section{Alfv\'en Mach number and gas and magnetic pressures}
 \label{Sec:quantitatives}
 
In Sect.~\ref{Sec:gradual}, we calculated magnetic and gas pressures, as well as Alfv\'en Mach numbers, to elucidate the step-by-step evolution of the field lines that resulted in the arc-shaped morphology shown in Fig.~\ref{fig:3DOrionField}. To estimate the magnetic and gas pressures, we used the following equations,

\begin{equation}
\begin{aligned}
&P_B = \frac{B^2}{8 \pi} [cgs],\\
&P_{\text{gas}} = nk_bT,
\end{aligned}
\end{equation}
where $P_B$, $B$, $P_{\text{gas}}$, $n$, $k_B$, and $T$ are the magnetic pressure, total strength of magnetic field, gas pressure, particle volume density, Boltzmann constant, and temperature, respectively. 

To determine the Alfv\'en mach number, we used the following equation:
\begin{equation}
\mathcal{M}_A = \frac{\sigma_{\varv}}{\varv_A},
\end{equation}
where $\sigma_{\varv}$ is the non-thermal velocity dispersion and $\varv_A$ is the Alfv\'en wave group velocity. The Alfv\'en velocity can be obtained using
\begin{equation}
\varv_A = \frac{B}{\sqrt{4 \pi \rho}} [cgs],
\end{equation}
where B is the strength of magnetic field and $\rho$ is the volume density.  Simulations  by~\citet{LiKlein2019} suggest that in the presence of strong fields ($\mathcal{M}_A \simeq 1 $), magnetic field lines can coherently bend around a formed filamentary molecular cloud as a result of converging flows, whereas for $\mathcal{M}_A \simeq 10$, complete distortion of magnetic fields is expected.

\end{document}